\documentclass[a4paper,12pt]{article}
\usepackage[paper=letterpaper,margin=1in]{geometry}
\usepackage{pifont}

\usepackage{amsmath,amssymb,amsfonts,epsfig,cite,setspace,bigstrut}
\begin{document}



\vskip 0.25in

\newcommand{\todo}[1]{{\bf ?????!!!! #1 ?????!!!!}\marginpar{$\Longleftarrow$}}
\newcommand{\fref}[1]{Figure~\ref{#1}}
\newcommand{\tref}[1]{Table~\ref{#1}}
\newcommand{\sref}[1]{\S~\ref{#1}}
\newcommand{\nn}{\nonumber}
\newcommand{\tr}{\mathop{\rm Tr}}
\newcommand{\comment}[1]{}

\newcommand{\cM}{{\cal M}}
\newcommand{\cW}{{\cal W}}
\newcommand{\cN}{{\cal N}}
\newcommand{\cH}{{\cal H}}
\newcommand{\cK}{{\cal K}}
\newcommand{\cZ}{{\cal Z}}
\newcommand{\cO}{{\cal O}}
\newcommand{\cB}{{\cal B}}
\newcommand{\cC}{{\cal C}}
\newcommand{\cD}{{\cal D}}
\newcommand{\cE}{{\cal E}}
\newcommand{\cF}{{\cal F}}
\newcommand{\cX}{{\cal X}}
\newcommand{\IA}{\mathbb{A}}
\newcommand{\IP}{\mathbb{P}}
\newcommand{\IQ}{\mathbb{Q}}
\newcommand{\IH}{\mathbb{H}}
\newcommand{\IR}{\mathbb{R}}
\newcommand{\IC}{\mathbb{C}}
\newcommand{\IF}{\mathbb{F}}
\newcommand{\IV}{\mathbb{V}}
\newcommand{\II}{\mathbb{I}}
\newcommand{\IZ}{\mathbb{Z}}
\newcommand{\re}{{\rm Re}}
\newcommand{\im}{{\rm Im}}
\newcommand{\li}{{\rm Li}}

\newcommand{\CA}{\mathbb A}
\newcommand{\CP}{\mathbb P}
\newcommand{\tmat}[1]{{\tiny \left(\begin{matrix} #1 \end{matrix}\right)}}
\newcommand{\mat}[1]{\left(\begin{matrix} #1 \end{matrix}\right)}
\newcommand{\diff}[2]{\frac{\partial #1}{\partial #2}}
\newcommand{\gen}[1]{\langle #1 \rangle}

\newcommand{\drawsquare}[2]{\hbox{%
\rule{#2pt}{#1pt}\hskip-#2pt
\rule{#1pt}{#2pt}\hskip-#1pt
\rule[#1pt]{#1pt}{#2pt}}\rule[#1pt]{#2pt}{#2pt}\hskip-#2pt
\rule{#2pt}{#1pt}}
\newcommand{\fund}{\raisebox{-.5pt}{\drawsquare{6.5}{0.4}}}
\newcommand{\antifund}{\overline{\fund}}

\newtheorem{theorem}{\bf THEOREM}
\def\thetheorem{\thesection.\arabic{theorem}}
\newtheorem{proposition}{\bf PROPOSITION}
\def\thetheorem{\thesection.\arabic{proposition}}
\newtheorem{observation}{\bf OBSERVATION}
\def\thetheorem{\thesection.\arabic{observation}}

\def\theequation{\thesection.\arabic{equation}}
\newcommand{\setall}{\setcounter{equation}{0}
        \setcounter{theorem}{0}}
\newcommand{\setequation}{\setcounter{equation}{0}}
\renewcommand{\thefootnote}{\fnsymbol{footnote}}

~\\
\vskip 1cm

\centerline{{\Large \bf Bipartita: Physics, Geometry \& Number Theory}}
\medskip

\vspace{.4cm}

\begin{center}
{\large Yang-Hui He}\\
\vspace*{3.0ex}
{\it
{\small
{Department of Mathematics, City University, London,\\
Northampton Square, London EC1V 0HB, UK;\\
School of Physics, NanKai University, Tianjin, 300071, P.R.~China;\\
Merton College, University of Oxford, OX14JD, UK
\footnote{Email: hey@maths.ox.ac.uk}
}
}}
\end{center}

\vspace*{4.0ex}
\centerline{\textbf{Abstract}} \bigskip
Bipartite graphs, especially drawn on Riemann surfaces, have of late assumed an active r\^ole in theoretical physics, ranging from MHV scattering amplitudes to brane tilings, from dimer models and topological strings to toric AdS/CFT, from matrix models to dessins d'enfants in gauge theory.
Here, we take a brief and casual promenade in the realm of brane tilings, quiver SUSY gauge theories and dessins, serving as a rapid introduction to the reader.

\vspace{1in}

{\it
Invited contribution to ``The Proceedings of The XXIX-th International Colloquium on Group-Theoretical Methods in Physics''
}
\newpage

\tableofcontents

\section{Observatio Curiosa}\label{s:obs}
We beg the readers to first indulge us with a seeming mere curiosity, and point out a peculiarity on a few theories well familiar to them.
Take perhaps the most famous supersymmetric field theory: $\cN=4$ super-Yang-Mills theory in four dimensions.
It is well known that this theory has three adjoint fields, say $X,Y,Z$, charged under a $U(N)$ gauge group, interacting via a superpotential which in $\cN=1$ language can be written as $W = \tr(X[Y,Z])$.
We could represent this data in terms of a so-called ``clover quiver'':
\begin{equation}\label{c3}
\begin{array}{ccc}
\begin{array}{l}\psfig{file=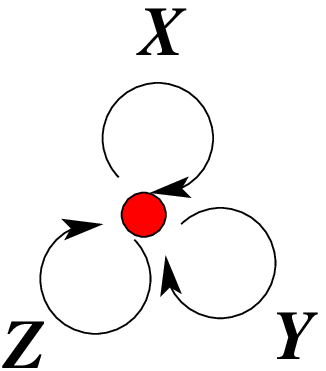,width=0.7in}\end{array}
& \qquad \qquad \qquad &
W = \tr(XYZ - XZY) \ .
\end{array}
\end{equation}
Here, as is customary, the nodes correspond to factors in the gauge group which generically will take the form $\prod_i SU(N_i)$ in the low energy and arrows correspond to fields charged appropriately as adjoints or bi-fundamentals.

We see that the number $N_G$ of gauge factors is one, the number of fields $N_F$, three and the number $N_W$ of terms in the superpotential, two. 
It is immediately evident that $1-3+2=0$.

To see that this is not a mere exception, let us move on to another famous super-conformal field theory in four dimensions, the so-called ``Klebanov-Witten conifold theory''.
This is an $SU(N) \times SU(N)$ theory with four fields $A_{i=1,2}$ and $B_{i=1,2}$ bi-fundamentally charged under the two gauge factors, interacting according to a quartic superpotential $W = \tr(\epsilon_{i\ell} \epsilon_{jk} A_i B_j A_\ell B_k)$ where $\epsilon_{ij}$ is the Levi-Civita symbol.
We can once again summarize this into a quiver, with the two nodes denoting the two $SU(N)$ gauge groups and the four arrows, the bi-fundamentals:
\begin{equation}\label{conifold}
\begin{array}{ccc}
\begin{array}{l}\psfig{file=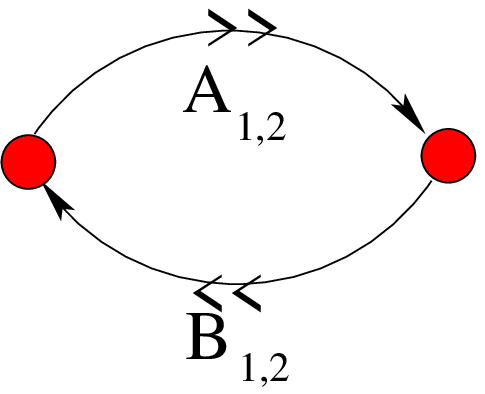,width=1in}\end{array}
& \qquad \qquad \qquad &
\begin{array}{l}
\begin{array}{ccc}
         & SU(N) & SU(N) \\
         A_{i=1,2} & \fund & \antifund \\
         B_{j=1,2} & \antifund & \fund
\end{array}\\
~\\
W = \tr(\epsilon_{il} \epsilon_{jk} A_i B_j A_l B_k)
\end{array}
\end{array}
\end{equation}
Once more we find that the number of gauge groups, here $N_g=2$, the number of fields, here $N_f = 4$ and the number of monomial terms in the superpotential, here $N_W=2$, satisfy the constraint that $2 - 4 + 2 = 0$.

Of the plenitude of $\cN=1$ supersymmetric gauge theories in four dimensions, especially the wealth thereof being engineered by the AdS/CFT correspondence as well as others geometrically attainable as low-energy limits in string theory, we can readily verify that the charming relation
\begin{equation}\label{T2rel}
N_g - N_f + N_W = 0
\end{equation}
persists for the majority of those in the literature.
This is not a coincidence: whenever the moduli space of the gauge theory is a toric variety, we shall see that \eqref{T2rel} holds.
In the context of string theory, almost all known gauge theories in the AdS/CFT correspondence have dual geometries which are toric because there are enough isometries for such spaces to allow for the explicit writing of (non-compact) Calabi-Yau metrics \cite{Gauntlett:2004hh}.
Indeed, our two examples above have moduli space of vacua being respectively $\IC^3$ and the conifold (a quadric hypersurface in $\IC^4$) which are canonical examples of affine toric varieties.
To further proceed with this point, it is expedient to briefly remind the reader of the concept of the vacuum moduli space (VMS) for a supersymmetric gauge theory, both within and independent of the string theory context.

\subsection{The Geometry of Moduli Space of Vacua}
Consider a generic supersymmetric gauge theory in four dimension with action, written in $\cN=1$ superspace notation, as given:
\begin{equation}
\int d^4x\ [ \int d^4\theta\ \Phi_i^\dagger e^V \Phi_i + 
\frac{1}{4g^2} 
\int d^2\theta\ \tr{\cW_\alpha \cW^\alpha} + 
  \int d^2\theta\ W(\Phi) + {\rm h.c.}]\ .
\end{equation}
Here $\Phi_i$ are chiral superfields, charged under some appropriate gauge group (not necessarily semi-simple), to which is associated some vector superfield $V$, field strength $\cW$ and coupling $g$;
the superpotential, $W$, is some (holomorphic) polynomial in $\Phi_i$.
In terms of the scalar component $\phi_i$ of $\Phi_i$, the effective potential is $\sum_i \left| \diff{W}{\phi_i} \right|^2 + \frac{g^2}{4}(\sum_i q_i |\phi_i|^2)^2$, with $q_i$ being the appropriate charges.
Thus, the VMS is the minimum of this potential, which, being non-negative, compels it to be the simultaneous solutions of
\begin{equation}\label{DF}
\mbox{F-Terms: } \diff{W}{\phi_i} = 0 \ ;
\qquad  \qquad
\mbox{D-Terms: } \sum\limits_i q_i |\phi_i|^2 = 0 \ .
\end{equation}
In general, these equations can become rather complicated, and the VMS, a non-trivial affine algebraic variety.
Utilization of algorithmic and computational geometry to systematically study these equations was recently initiated \cite{Gray:2006jb}.

In string theory, when the gauge theory is engineered as the world-volume physics of a stack of D3-branes, the VMS is by construction the transverse six-dimensions which the branes probe.
This is the key to the algebraic geometry of the AdS$_5$/CFT$_4$ correspondence: the VMS of world-volume theories is some affine Calabi-Yau threefold, existing as a real cone over a Sasaki-Einstein 5-manifold.
Our two above examples, \eqref{c3} and \eqref{conifold}, are both embeddable into string theory, and we can readily check that the VMS are, respectively, (symmetrized products of) $\IC^3$ as a cone over $S^5$ and the conifold as a cone over a space known as $T^{1,1}$. 

By far the largest class of geometries which has dominated the literature, especially in AdS/CFT, is toric varieties and on these we shall henceforth focus.
Moreover, for simplicity, we will take all the nodes of our quivers to be labeled by $N=1$ so that our gauge theories are Abelian; one could easily promote to arbitrary $N$, resulting in the VMS being an $N$-th symmetrized product.
It shall be too great a digression for us to review toric geometry here, we simply mention that when the VMS is toric, one could use the gauge linear sigma-model \cite{Witten:1993yc} to study how the brane resolves singularities \cite{Douglas:1997de}.
The algorithm of engineering the gauge theory, given an arbitrary toric Calabi-Yau diagram, was traditionally done by partial resolutions \cite{Feng:2000mi}.

\section{Toric Moduli Space, Gauge Theory, Bipartiteness, etc.}
\begin{figure}[!h!t!b]
\centerline{\psfig{file=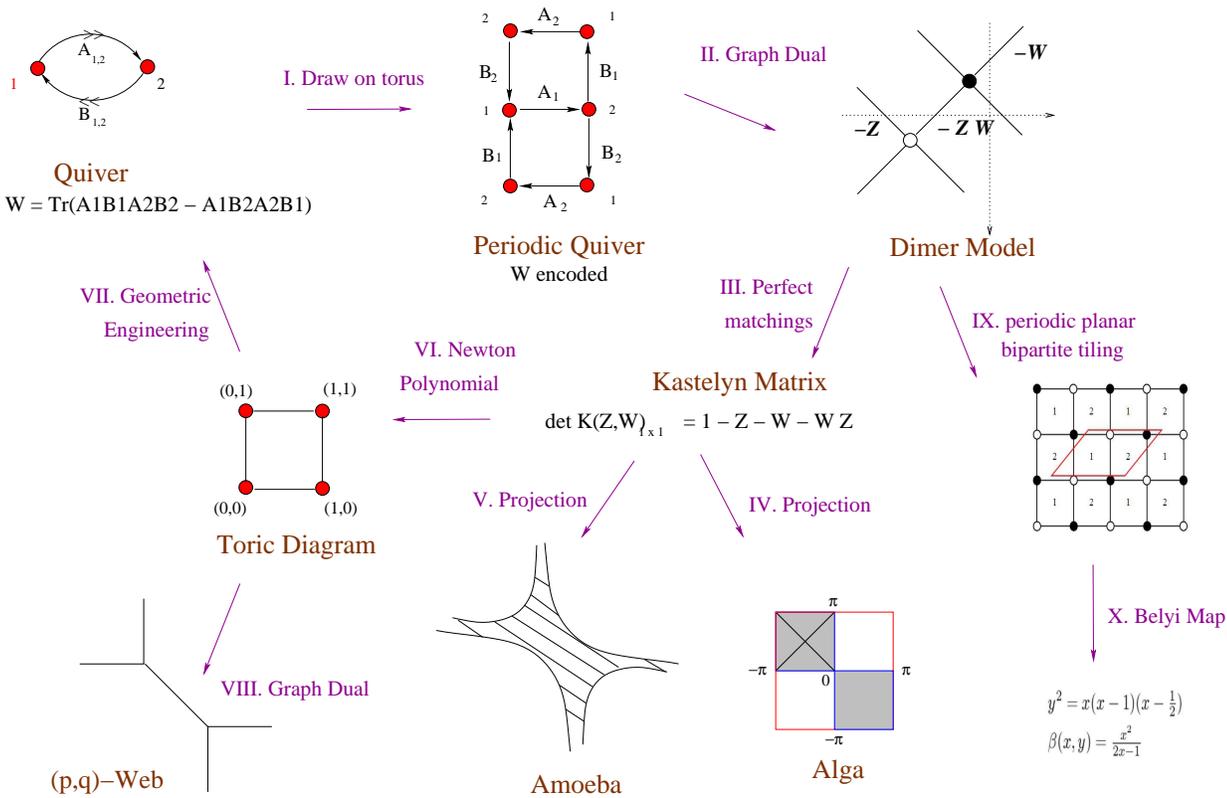,width=7in}}
\caption{{\sf \em The intricate web of correspondences, illustrated for the conifold theory, ranging from quiver diagrams to bipartite graphs, from $(p,q)$-webs to brane tilings and from amoeba projections to Belyi maps.}}
\label{f:web}
\end{figure}

The astute reader would recognize \eqref{T2rel} as the Euler relation for a torus and we shall see how the aforementioned gauge theories can be written as tilings thereon, this observation thus initiated this enormous subject of dimer models and brane tilings \cite{Hanany:2005ve}.
The limitation of space does not permit us to delve into the vast web of correspondences which has emerged over the past decade and one is referred to marvelous reviews \cite{Kennaway:2007tq,Yamazaki:2008bt} and the myriad of works in both the physics and mathematics communities from which we draw a small sample representative of the various directions \cite{Franco:2005rj,GarciaEtxebarria:2006aq,Butti:2006au,Feng:2005gw,stienstra,Ashok:2006br,Koch:2010zza,Jejjala:2010vb,Hanany:2011ra,Hanany:2011bs,He:2012xw,broomhead,Franco:2012mm,He:2012kw}.
What we will do is to walk through the Figure \ref{f:web}, which illustrates the case for the conifold theory in some detail, and hope an appreciation, in a rather concrete manner, could be inspired in the reader.

\paragraph{Arrow I: }
We begin with the theory, already presented in \eqref{conifold}, having the quiver with four arrows and the quartic superpotential $W$.
Following Arrow I, we recognize that if we drew the quiver periodically on the plane (with appropriately identified nodes and arrow explicitly marked), then, in the fundamental region, there will be two loops, one counter-clockwise, and the other, clockwise; these are precisely the two terms in the superpotential $W$.

\paragraph{Arrows II \& III: }
The graph dual - by replacing a counter-clockwise loop with a black node and clockwise, white - is then a bipartite graph periodic on a plane; this is more apparent when zooming out a little following Arrow IX.
For the conifold, this is a rectangular tiling with two squares in the fundamental domain $\cF$.
Equivalently, we can draw this as a {\em dimer model} on a torus -- the origin of \eqref{T2rel}.
Marking the two cycles of the $T^2$ with formal variables $z$ and $w$ and the (signed) crossings of the dimer relative to these cycles in $\cF$ produce the indicated monomials.

\paragraph{Arrow IV: }
In statistical mechanics and chemistry, an important problem is the enumeration of the number of {\em perfect-matchings} (where each pair of white-black is linked by a bi-molecular bond).
This is achieved by a weighted adjacency, called the {\em Kasteleyn matrix} (here, there is only one pair of black-white nodes, thus it is just $1 \times 1$), whose determinant $K$ gives a bi-variate polynomial in $(z,w)$.

\paragraph{Arrows V \& VI: }
Treating $K(z,w)$ as an algebraic variety (complex curve), we can perform two projections:
\begin{enumerate}
\item the {\em amoeba} projection $(z,w) \rightarrow (\log|z|,\log|w|)$ and 
\item the {\em alga} (or co-amoeba) projection $(z,w) \rightarrow (\arg(z),\arg(w))$.
\end{enumerate}
These are real algebraic sets which constitute the study of ``tropical geometry''.
One can see that the alga projection, being itself doubly periodic (defined, for instance, in the fundamental region $[-\pi,\pi]^2$), very nicely contracts to the dimer.

\paragraph{Arrow VII: }
The curve $K(z,w)$ is the {\em Newton polynomial} for the {\em toric diagram} of the conifold. 
This is a standard book-keeping device in polytope theory: for a set of planar lattice points $D = \{(a_i,b_i)\}$, the Newton polynomial encodes $D$ by
$K_D(z,w) := \sum\limits_D z^{a_i}b^{b_i}$ (of course, this is in general a Laurent polynomial because $(a_i,b_i)$ may be negative integers).
Indeed, non-compact toric Calabi-Yau threefolds have the property that their toric diagram is a set of co-planar lattice points.
Remarkably, the mirror of a non-compact toric Calabi-Yau threefold with toric diagram $D$ is given by the hypersurface $u v = K_D(z,w)$ in $\IC^4[u,v,w,z]$.
Thus, we can explicitly map the various inter-connections, by thrice T-duality, to the mirror, whereby giving us a parallel set of correspondences.

Given a geometry, such as specified by the toric diagram, the inverse algorithm of extracting the physics of the quiver theory is the process of geometrical engineering.
The reverse of this arrow, of obtaining the geometry from the physics, is the process of computing the VMS to which we alluded in \eqref{DF}.
In some sense, the investigation of this arrow was the original motivation and the incipience of this rather unexpected web of inter-relations.

\paragraph{Arrow VIII: }
Whereas by Arrow IV, the alga contracts to the dimer, 
the complementary amoeba projection, contracts nicely to the graph-dual of the (resolution) of the toric diagram.
This dual is called the {\em $(p,q)$-web} and is where charged branes can wrap collapsing cycles. 
In summary, the $(p,q)$-web is the ``spine'' of the amoeba of $K(z,w)$ while the dimer is the contraction of the alga of $K(z,w)$.

\paragraph{Arrow X: }
When delving into bipartite graphs drawn on Riemann surfaces, one cannot resist but enter the subject of {\em dessins d'enfants}, which, as outlined in his ``Esquisse d'un Programme'', impressed Grothendieck more than any other topic in mathematics.
For any such graph, one could find a Riemann surface $\Sigma$ defined over the algebraic numbers $\overline{\IQ}$ such that a rational map $\beta$ takes $\Sigma$ to $\IP^1$ ramified only at three points: $0,1,\infty$.
The data $(\Sigma, \beta)$ is called a {\em Belyi pair}.
It is an extraordinary theorem of Belyi that this is a necessary and sufficient condition, that is, $\Sigma$ has a model as an algebraic curve with coefficients in $\overline{\IQ}$ if and only if there exists such a map $\beta$.

For us, $\Sigma$ is the torus realized concretely as a Weierstra\ss\ equation of an elliptic curve, which, by Belyi, can be defined over $\overline{\IQ}$.
The pre-images of 0 (respectively 1) are the black (resp.~white) nodes whose valency is equal to the ramification index of $\beta$, and pre-images of $\infty$ are the faces.
In this way, our gauge theory is completely encoded by a dessin on an elliptic curve, which is the pre-image $\beta^{-1}$ of an interval $[0,1]$ on $\IP^1$. 
We can easily check that for our example, $\beta^{-1}(0)$ is the point $(0,0)$ on $y^2 = x(x-1)(x-\frac12)$ at which the map vanishes to order 4, corresponding to quadrivalent black node;
$\beta^{-1}(1) = (1,0)$, also of ramification order 4, corresponding to the quadrivalent white node;
and finally $\beta^{-1}(\infty)$ has two points $(\frac12,0)$ and $(\infty,\infty)$, each of ramification index 2, corresponding to the two square faces.

\section{Prospectus and Open Problems}
We have only taken a glimpse of the extensive geography of a {\it terra Carmeli}; there is certainly much to explore.
A rich interplay between physics and mathematics is woven into the fabric of this web.
For example, Seiberg duality translates to so-called quiver mutations (or, in the mirror, Picard-Lefschetz monodromy actions), which in turn becomes ``urban-renewal'' moves in the bipartite graph.
These are profound geometrical realizations of duality in field theory.

Moreover, there are many interesting manifestations of number theoretic properties as well.
For example, it was realized that R-charges in super-conformal field theories in the toric AdS/CFT context are algebraic numbers \cite{Martelli:2005tp}, this ties in nicely with the fact that all our bipartite graphs are dessins embedded on elliptic curves with algebraic coefficients and the fact that the degree of extension over $\IQ$ of the R-charges is a Seiberg duality invariant \cite{Hanany:2011bs}.

With the emergence of elliptic curves, it is natural to compute the Klein $j$-invariant.
As a parting digestif, let us leave the reader with a puzzle which has so far eluded analysis \cite{He:2012xw}.
There are three tori here: (1) the torus on which the dimer is drawn, most naturally with the length determined by R-charges in a so-called isoradial embedding,
(2) the elliptic curve on which the dessin is drawn and (3) the torus inside the $T^3$ of mirror symmetry (constituting our Calabi-Yau threefold as the fibre over special Lagrangian cycles).
When are these isogenous, as detected by the equality of the respective $j$-invariants?
For simple geometries such as quotients of $\IC^3$ or the conifold, all three are isogenous. However, for more complicated examples, the isoradial embedding does not seem the one which matches the dessin while, perhaps more intriguingly, the $j$-invariants of (2) and (3) differ by an inexplicable minutia (cf.\S6.3 of \cite{He:2012xw}).

~\\

\centerline{ \reflectbox{\ding{167}}---\ding{69}---\ding{167} }

\section*{Acknowledgments}
{\it Ad Catharinae Sanctae Alexandriae et Ad Majorem Dei Gloriam.}\\
I would like to thank the organizers of ``The XXIX International Colloquium on Group-Theoretical Methods in Physics'' for inviting me to the wonderful conference and take this opportunity to express my gratitude to my collaborators N.~Benishti, S.~Benvenuti, B.~Feng, D.~Forcella, S.~Franco, D.~Galloni, A.~Hanany, V.~Jejjala, K.~Kennaway, J.~McKay, J.~Pasukonis, S.~Ramgoolam, D.~Rodriguez-Gomez, J.~Sparks, A.~Uranga, C.~Vafa and A.~Zaffaroni for our many adventures in {\it terram bipartitam.} To the STFC, UK, for an AF and for grant ST/J00037X/1, to the Chinese Ministry of Education, for a Chang-Jiang Chair Professorship at NanKai University, and to the US NSF for grant CCF-1048082 I am most indebted.


\end{document}